\documentclass[two column,prl]{revtex4}
\usepackage{bm}
\usepackage{graphics}
\usepackage{bm}
\usepackage{graphics}
\usepackage{graphicx}
\usepackage{amsfonts}
\usepackage{amsmath}
\usepackage{amssymb}
\usepackage{graphicx}
\usepackage{color}
\usepackage{fancyhdr}

\begin{document}

\preprint{HEP/123-qed}
\title[Short title for running header]{\texttt{\texttt{Electron Emission Projection Imager}}}
\author{Stanislav S. Baturin\footnote{Current address: Department of
Physics, The University of Chicago, 5720 S. Ellis Ave., Chicago,
IL 60637, USA}}
\author{Sergey V. Baryshev}
\email{sergey.v.baryshev@gmail.com} \affiliation{Euclid TechLabs,
365 Remington Blvd., Bolingbrook, IL 60440, USA}

\begin{abstract}
\noindent A new projection type imaging system is presented. The
system can directly image the field emission site distribution on
a cathode surface by making use of anode screens in the standard
parallel plate configuration. The lateral spatial resolution of
the imager is between 1 and 10 $\mu$m. The imaging sensitivity to
the field emission current can be better than the current
sensitivity of a typical electrometer, i.e. less than 1 nA.
\end{abstract}

\maketitle

\pagenumbering{gobble}

\noindent \textbf{Introduction}

\emph{In situ}, real-time or time-resolved imaging at micro- and
nano-scales has provided a great deal of understanding of various
processes that take place in materials. Many contemporary imaging
concepts are intrinsically based on raster scanning. This means
that the pointed focused/sharp probe, be it an electron or ion
probe, or a mechanical cantilever, scans across a given area to
resolve and plot an image of a static or dynamic pattern. Such
images consist of pixels or voxels containing locally collected
information. Examples are SEM, STEM, STM, AFM, imaging TOF SIMS,
and others.

A second approach is much older and comes from photography. This
is the projection method and it is entirely different from raster
scanning. In this approach, a surface of interest under
electromagnetic irradiation or electron bombardment is imaged at a
certain distance beyond the interaction location giving rise to a
significant magnification. Examples are electron diffraction in
TEM and LEED, and X-ray topography. The most remarkable examples
here would be the field emission microscope and field ion
microscope invented by Erwin Mueller, in which electron or ion
imaging of a sharp tip placed in a high electric field becomes
possible with nanometer or atomic resolution without complex
motorized systems, data acquisition, and post-processing methods.

In the realm of finding inexpensive, reliable and adaptive field
emission sources, beyond Spindt cathodes, to drive various devices
and systems, many novel advanced materials have been studied. This
research field remains extremely active and is of significant
interest to this very day. Field emission sources often yield
laterally non-uniform emission. This is because field emission is
a non-linear process and therefore surface termination and
geometrical features contribute significantly to the surface
barrier formation and its lateral uniformity. Thus, in practice a
packaged field emission source will always have strong and weak
emission points.

To the best of our knowledge, systems to observe the lateral
uniformity of field emission arrays are always raster-scanning
tools that move a micron scale anode tip across a field emitter
surface \cite{1,2,3}. In this paper, we present an electron
emission projection imager system that captures lateral emitter
distribution of the entire surface instantaneously.

\

\noindent \textbf{General description of the imager. Imaging and
measurement examples}

The conceptual diagram of the imaging system and its actual
appearance are shown in Fig.1. The imager has three anodes (1) an
optically polished (1 inch dia. and 100 $\mu$m thick) disk made of
yttrium aluminum garnet doped with cerium (YAG:Ce) which is coated
with a Mo film of about 7-8 nm in thickness; (2) a transparent and
conductive tin-doped indium oxide (ITO) film of 175 nm deposited
onto a boro-aluminosilicate glass (BASG) substrate of 700 $\mu$m
in thickness; (3) a polished (1 inch dia. and 100 $\mu$m thick)
disk made of stainless steel (SS).

The Mo/YAG:Ce and ITO/BASG anodes are imaging screens and the SS
anode is to measure long-term temporal current stability.
Mo/YAG:Ce and ITO/BASG are semitransparent in the visible range.
Fig.2a illustrates detailed UV-vis. spectra and Fig.2b
demonstrates transparency when anode screens covering color lines
are backside illuminated.

Depending on the specific measurement to be performed, the anodes
can be interchanged by linearly moving the frame holding all three
anodes, by means of an ultrahigh vacuum actuator. The distance
between the sample and the anode is set using a micrometer holding
the sample. Top and side view cameras outside the vacuum are used
to check the parallelism between the cathode and anode, and to
measure the gap. A Canon DLSR camera is installed at a viewport
behind the anodes to take pictures of electron emission patterns.
The sample electrode is at ground and the anode frame is isolated
and positively biased. The bias and current readings are enabled
by a Keithley 2410 electrometer. The Keithley and the Canon are
programmed such that they collect data sets of current-voltage
(I-V) curves and images synchronously.

\begin{figure*}[t]
\includegraphics[width=11cm]{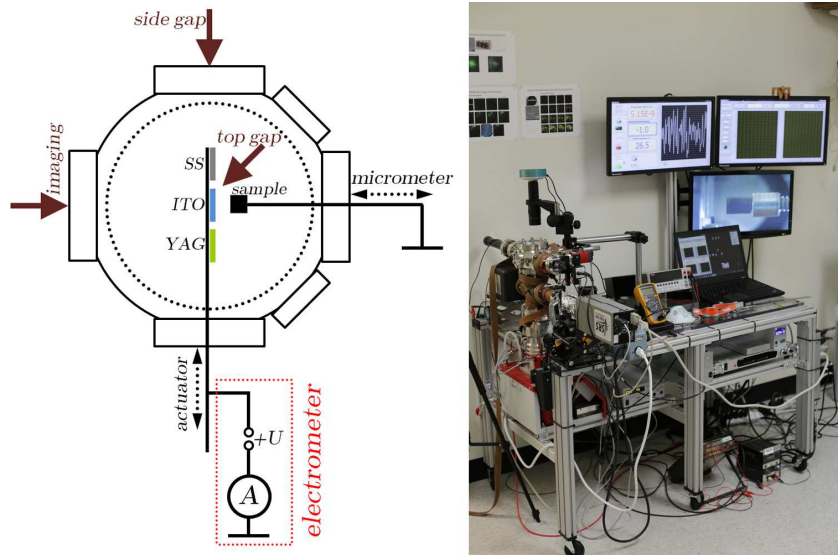}

\caption{A schematic of the electron emission imager. Solid arrows
represent the 3 cameras: two cameras are used to monitor
parallelism between the cathode and anode surfaces upon
installation and to evaluate the interelectrode gap during field
emission measurements, and the third camera is used to detect
light from either anode screen luminescence or light originating
from the cathode surface itself (exampled below), or both}
\end{figure*}

\begin{figure*}[t]
\includegraphics[width=11cm]{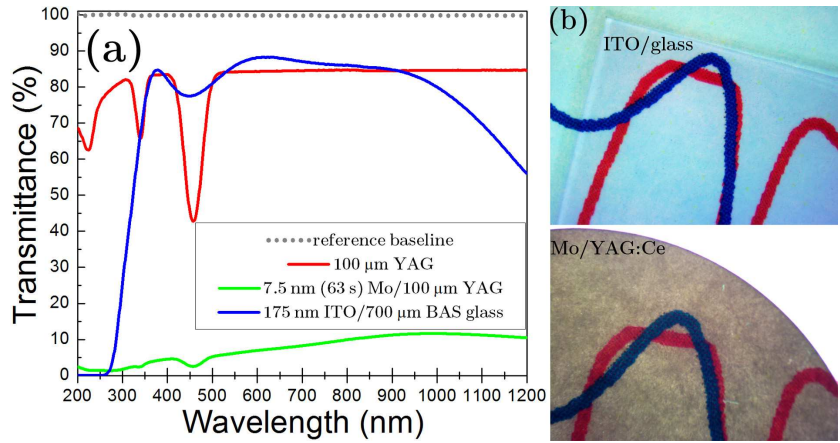}

\caption{(a) UV-vis. spectra (spectral dependence of
transmittance) for the YAG:Ce (red line), Mo/YAG:Ce (green line)
and ITO/BASG (blue line) anode screens. (b) Photographs that
demonstrate the semitransparency of the ITO/BASG anode (top) and
Mo/YAG:Ce anode (bottom)}
\end{figure*}

Before going into a detailed description of the imager, we would
like to introduce a series of pictures taken on a carbon nanotube
(CNT) sample grown by a CVD process. Fig.3a illustrates the
appearance of the sample in an optical micrograph, Fig.3b
illustrates the morphology as measured using a scanning electron
microscope (SEM), Fig.3c illustrates the green (550 nm) light
produced by the Mo/YAG:Ce anode screen under $\sim$1 keV and
$\sim$100 $\mu$A electron bombardment, and Fig.3d illustrates
clusters of intense red light imaged with the ITO/BASG anode
screen under similar $\sim$1 keV and $\sim$100 $\mu$A conditions.
The red cores already can be seen in Fig.3c, because the red light
intensity is extremely high and the Mo/YAG:Ce screen is
semitransparent in the red. The red light thermally produced by
CNT due to electric current flow induced heating is a well-known
effect which has been documented over the last 15 years
\cite{4,5,6,7,8}. Note -- the Mo/YAG:Ce and ITO/BASG anode images
were taken at different sample installations and different camera
presets and therefore the red light patterns do not coincide.

\begin{figure}[t]
\includegraphics[height=8cm]{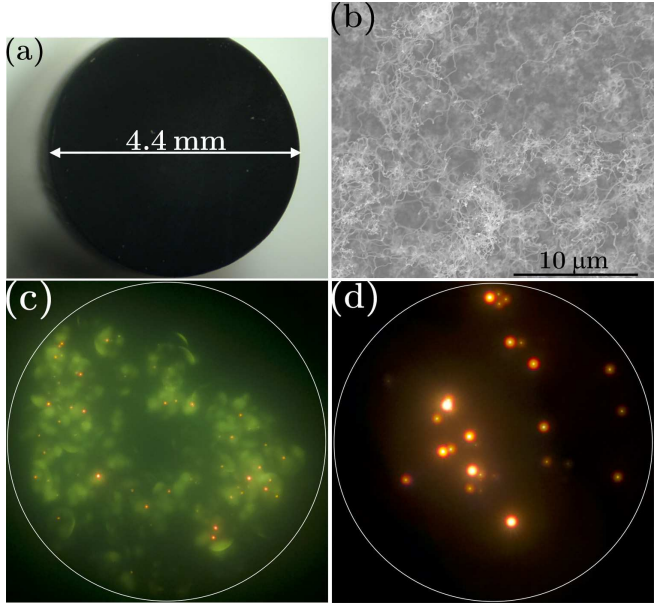}

\caption{CNT on stainless steel substrate: (a) an optical
microscopy image of the CNT sample; (b) a high magnification SEM
image showing the CNT morphology; (c) emission pattern imaged with
the Mo/YAG:Ce anode screen; (d) emission pattern imaged with the
ITO/BASG anode screen. The field of view (the white circle
diameter) in the fragments (c) and (d) is 4.4 mm}
\end{figure}

Along with imaging, I-V curves can be collected as demonstrated in
Fig.4a. Linear and semi-logarithmic I-V characteristics of the CNT
sample are presented. The abscissa is presented in units of
electric field after the recorded voltage was divided by the
interelectrode gap taken by the top camera (Fig.4b, right). As
mentioned, the opaque SS anode is used to measure long-term
emission current stability. An example of the results of a long 24
hour run carried out for the CNT sample is shown in Fig.4b.

Spatial resolution of the acquired images is limited by the
optical system. For the case shown in Fig.3c, the resolution is 8
$\mu$m per pixel (the known sample diameter is 4,400 $\mu$m and
the apparent sample diameter in Fig.3c is 550 pixels). Using
another lens we recently were able to consecutively achieve higher
optical magnifications which resulted in higher spatial
resolutions of 5, 2 and 1 $\mu$m per pixel.

\begin{figure*}[t]
\includegraphics[width=13cm]{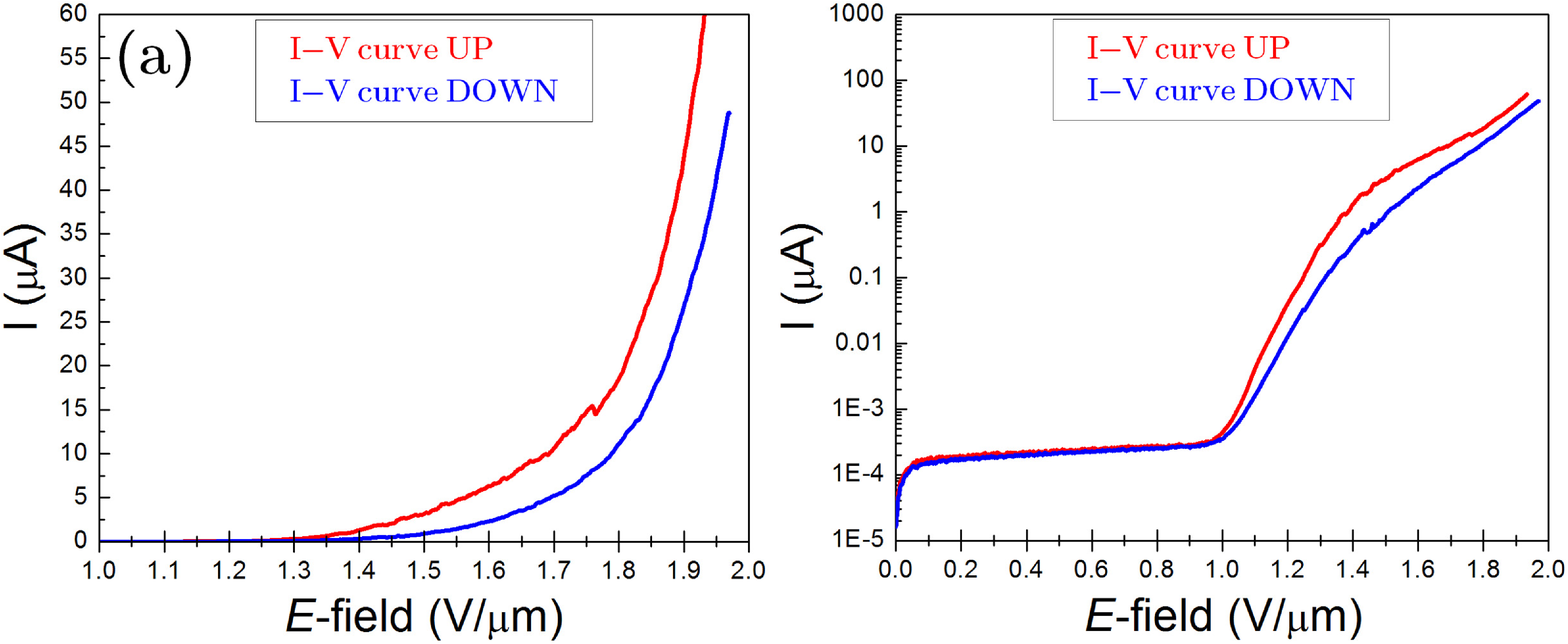}
\includegraphics[width=13cm]{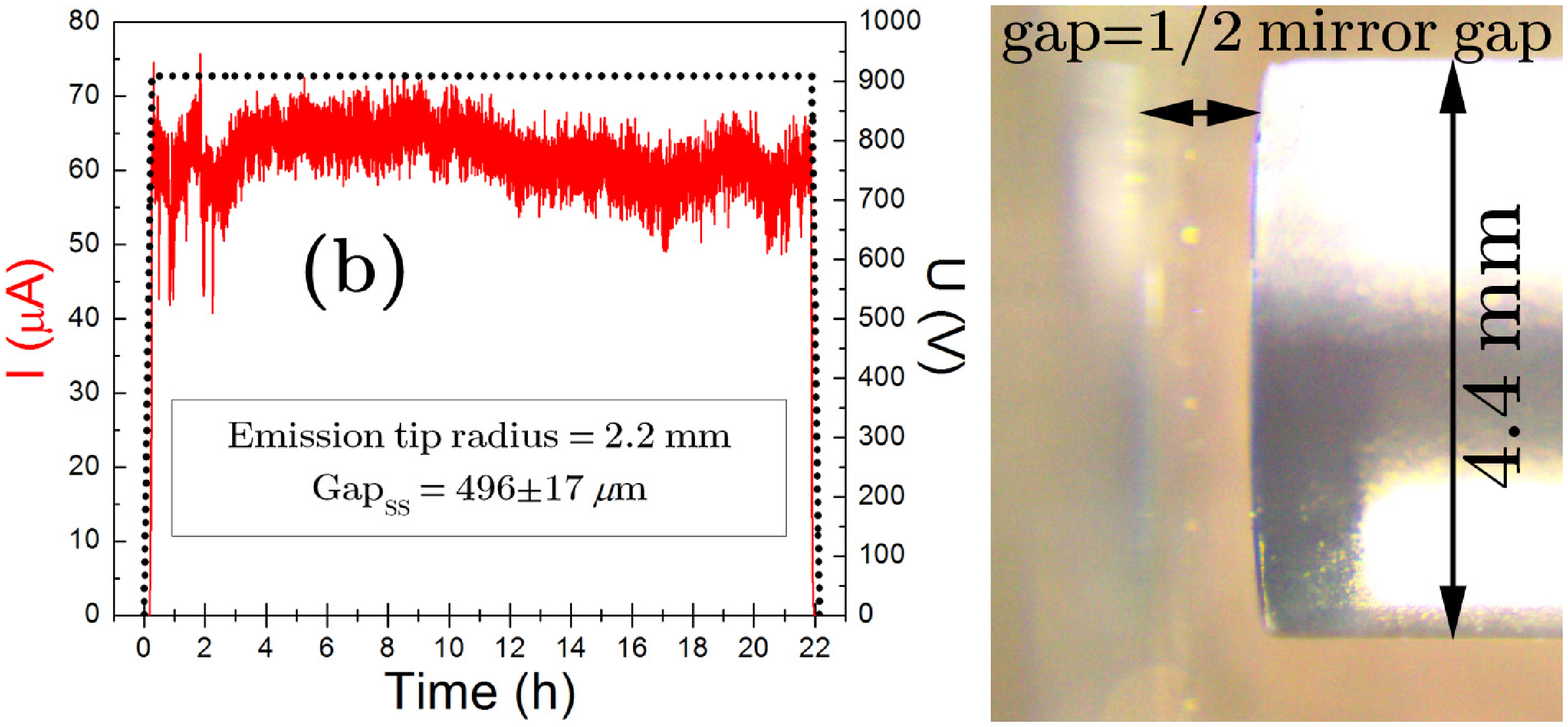}

\caption{Electric characterization of the CNT sample: (a) I-V
characteristics in linear and semi-log representation and (b)
current time stability measurement taken over a 24 hour period.
The right panel demonstrates an example of electrode parallelism
and the gap between the SS anode and the CNT cathode measured as
496 $\mu$m}
\end{figure*}

\newpage

\noindent \textbf{Detailed description of the apparatus}

\noindent \emph{Imaging screens}

ITO/BASG screens were purchased from Delta Technologies. The anode
screen is a 1 sq. inch BASG square, 0.7 mm in thickness. An ITO
film of approximately 175 nm is deposited on one side of the BASG
square. The ITO film resistance is between 4-10 $\Omega$. The
conductive ITO side faces the cathode under measurement
conditions.

To establish an electric field between the sample and the
isolating YAG:Ce anode and to collect the current, a metal film
needs to be deposited. Molybdenum was chosen because it is dense
and allows for continuous ultrathin films. In addition, the
melting point of Mo is 2,896 K, meaning that Mo coatings can
sustain exceptionally high electrical power surface densities
$\frac{I\cdot V}{cm^2}$. We specify this because originally Al
coatings were used that would melt upon the first run producing
bulged anode surfaces at the location where current was collected.
Mo is deposited in house by magnetron sputtering on one side of an
optical-quality polished YAG:Ce disk. Base pressure in the
magnetron system is $<$5$\times$10$^{-7}$ Torr. Prior to coating,
the YAG:Ce disk is cleaned \emph{in situ} using RF discharge
plasma. Without breaking vacuum, immediately after the cleaning,
the Mo coating is deposited. Ar is used as a working gas for both
cleaning and sputtering at a pressure $\sim$10$^{-3}$ Torr.

For voltages less than 1.1 kV, provided by the Keithley 2410, the
Mo coating has to be ultrathin. A standard deposition time of 60
seconds was found to be optimal for our application. This means
the thickness of the resulting Mo film is sufficiently small to
let many electrons into the YAG:Ce to produce photons, but thick
enough to filter electrons such that they do not penetrate too
deeply into the YAG:Ce phosphor -- this would lead to charging and
discharging effects in the YAG:Ce. For instance, a Mo thickness
corresponding to a 60 s deposition time does not comprehend
applied voltages over 2 kV. The electron penetration depth becomes
such that electrons start accumulating inside the YAG:Ce and do
not efficiently drain. This results in luminescence instabilities
and may lead to strong discharge events damaging both the YAG:Ce
and Mo coatings. The choice of Mo thickness is determined by the
application and the voltage range to be applied between the
electron emitter under study and the anode.

For specific and quantitative choice of the Mo thickness, the
Kanaya-Okayama approach \cite{9} can be used. The following
formula relates the electron energy, $E$ [keV], to the electron
range, $RKO$ [$\mu$m], the maximum possible penetration depth

\[
RKO=27.6\cdot 10^{-3}\cdot \frac{MM\cdot
E^{1.67}}{Z^{0.889}\cdot\rho} \text{, (1)}
\]
where $MM$ is the molar mass [g/mol], $Z$ is the atomic number,
$\rho$ is the mass density [g/cm$^3$]. The resulting dependence is
shown in Fig.5, and provides a precise answer for the maximum
thickness of Mo. The example calculation at 1 keV yields $RKO$=9.3
nm. In practice, one would use not the maximum range but rather
so-called practical range which is about 0.7 of $RKO$ \cite{10}.
This leads to a specific requirement on the Mo coating -- 6.5 nm.
At this Mo thickness, about 10\% of electrons that were not back
scattered, will penetrate into the YAG:Ce and produce photons.

\begin{figure}[t]
\includegraphics[width=8cm]{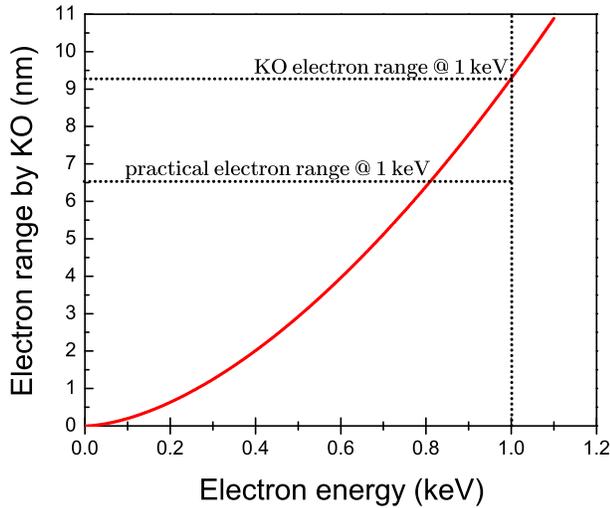}

\caption{The dependence of the electron range on the energy,
calculated for Mo using the KO formula, Eq.(1)}
\end{figure}

Now we can explain why the deposition time of 60 seconds works
best for the 1.1 keV application by converting the deposition time
into the actual thickness of the Mo coating. To measure the
thickness of the Mo coating, we synthesized three test samples on
optically polished Si witness coupons and depth profiled them
using secondary ion mass spectrometry (SIMS). One sample was
deposited for 60 s (the same time used for YAG:Ce anode
fabrication), and two others were deposited for 90 s and 120 s.
With SIMS, it is possible to measure the thickness. By definition,
sputtering rate ($SR$) is calculated by dividing the layer
thickness $d$ by the time $T$ it took to sputter through it. On
the other hand, the $SR$ can be expressed through other
experimental and fundamental material parameters: the ion current
density $j$, the sputtering yield (i.e., the number of ejected
atoms per incident ion) $Y$, the elementary charge $e$, and the
atomic density of the target material (units of atoms per volume)
$N$, or alternatively the mass density of the target material
$\rho$ and the atomic mass of the target material $M$. Combining
everything together, we can deduce a formula for the thickness of
the layer to be determined as

\[
\left\{
\begin{array}{ll}
SR=\frac{d}{T}\\
SR=\frac{j\cdot Y}{e\cdot N}\Longrightarrow d=T\cdot\frac{j\cdot Y\cdot M}{e\cdot\rho}\\
N=\frac{\rho}{M}
\end{array}
\right. \text{. (2)}
\]

The test samples were depth profiled using a time-of-flight SIMS
instrument in the traditional single beam mode \cite{11}. The
60$^\circ$ oblique 5 kV Ar$^+$ ion beam was focused to a spot of
30 $\mu$m in diameter at a current of 80$\pm$8 nA as measured by
\emph{in situ} Faraday cup. When depth profiling, the beam was
raster scanned over an area of 675$\times$675 $\mu$m$^2$, yielding
an ion current density $j=17.6\pm1.8$ $\mu$A/cm$^2$. Both the
single ion beam spot and the raster sizes were measured using the
\emph{in situ} optical Schwarzschild microscope that directly
images the surface of the sample and its modification in real time
\cite{12}. Electronic gating during the profiling experiments was
set to 50\%. The resulting depth profiles are illustrated in
Fig.6. We determine the time $T$ at which the Mo film is eroded
away as a moment at which the Si and Mo (main isotopes 28 and 98
atomic mass units respectively) signals cross. With this fixed,
the Mo film deposited for 60 s was eroded in 72 s, the one
deposited for 90 s was eroded in 109 s, and the one deposited for
120 s was eroded in 144 s. Using Eq.(2), three thicknesses 7.5 nm
(60 s deposition), 11.4 nm (90 s deposition) and 15.1 nm (120 s
deposition) were calculated with $j=17.6\pm1.8$ $\mu$A/cm$^2$;
$M$=95.95 a.m.u.=95.95$\times$1.7$\times$10$^{-24}$ g;
$\rho$=10.28 g/cm$^3$; $e$=1.6$\times$10$^{-19}$ C; $Y$=6.02.
$Y$=6.02 was obtained using SRIM/TRIM \cite{13}, a Monte-Carlo
code, in the full cascade mode on the ensemble of $10^5$ Ar$^+$
ions with their energy set to 5 keV and the irradiation angle set
to 60$^\circ$.

We stress that (1) the apparent spike for the Si signal is due to
the presence of the C-O molecule on the surface (no isotopic
pattern for Si was observed in a full mass spectrum recorded for
the pristine surface, sputtering time=0 s) and (2) the apparent
spike for the Mo signal is due to slight metal surface oxidation
that always leads to orders of magnitude higher secondary ion
signals in SIMS (secondary ion formation probability is higher in
metal-oxides compared to clean metals) \cite{14}.

The thickness of 7.5 nm explains why the standard Mo/YAG:Ce screen
is semitransparent in the visible range (Fig.2), and thus why we
observe the combination of green and red photons when imaging
electron emission with the YAG:Ce screen (Fig.3). The determined
7.5 nm thickness of the standard Mo coating, being in excellent
agreement with electron range calculations, straightforwardly
explains why deposition time of 60 s is optimal for imaging with
the YAG:Ce phosphor screen at applied voltages of up to 1.1 kV.

\begin{figure}[t]
\includegraphics[width=8cm]{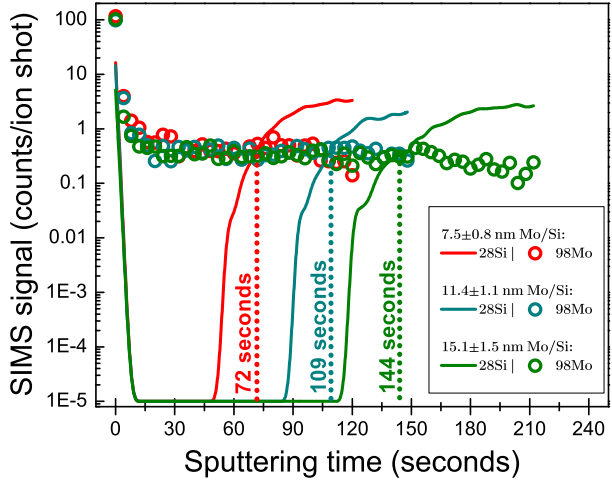}

\caption{Depth profiles for Mo films on Si witness coupons}
\end{figure}

\

\noindent \emph{Photon sensitivity}

Taking it a step forward, let us estimate photon sensitivity of
the system. The first step here would be to calculate the electron
to 550 nm photon conversion as

\[
C_{e\rightarrow ph}=T_e^{vac/Mo}\cdot T_e^{Mo}\cdot C_{YAG}
\text{, (3)}
\]
where $T_e^{vac/Mo}$ is the transmission coefficient of electrons
through the interface between vacuum and the Mo film (1 minus
reflection), $T_e^{Mo}$ is the transmission coefficient of
electrons through the Mo film, $C_{YAG}$ is electron-to-photon
conversion of all the electrons that penetrate through the Mo film
and get embedded in the YAG:Ce screen. $T_e^{vac/Mo}$ is only
dependent on $Z$ and is approximately equal to 0.7 \cite{10}.
$T_e^{Mo}$ is equal to 0.1 at a Mo thickness close to practical
electron range (6.5 nm at 1 keV). We consider $C_{YAG}$=0.01
\cite{15}.

Assuming that 1 keV electrons provide a relatively uniform
distribution of a current of 100 $\mu$A over an area of 2.2 mm in
radius (Fig.1) captured by the Canon 6D on the 500 by 500 pixels
sensor area and using the resulting $C_{e\rightarrow
ph}=7\times10^{-4}$, we obtain that the signal per pixel is
$\sim$$10^6$ photons/s with a Poisson noise of $\sim$$10^3$ and a
S/N ratio $\approx$$10^3$. If the current (emission is still
considered uniform) is reduced to 100 nA and further to 1 nA (the
limit for the Keithley), then the signal per pixel reduces to
$\sim$$10^3$ (S/N$\approx$30) and 10 (S/N$\approx$3) photons/s,
respectively. Since the read noise level at ISO=1,000 (used in
experiments) in the Canon 6D is between 1 to 5 electrons, such a
low electron emission should be still observable by the camera at
1 keV. This would be worst case scenario. In fact, we typically
see that emission starts with isolated small features, meaning
that even though the current is reduced by many orders of
magnitude the local pixel photon density on the sensor can remain
very high.

The main factor here is not the current or current density
providing the Canon sensor with enough photons per pixel, but the
electron energy that should remain higher than 700 eV in order to
keep the practical electron range no less than 5 nm. In this
sense, Mo/YAG:Ce anode screens with different Mo thicknesses can
be combined (up to 3 on the frame holder shown on the schematic
diagram in Fig.1) and/or a larger inter-electrode gap used to
enable higher electron kinetic energy at the same electric field
in order to detect small signals.

At 100 $\mu$A and 1 keV and ISO=1,000, we can calculate the
photographic equivalent of the light produced on the Mo/YAG:Ce
screen, namely, $\sim$$10^6$ photons per pixel per second
corresponds to $\sim$$10^{16}$ photons per second per m$^2$. At
the wavelength of 550 nm, this is about 1/2 lux or 1 EV. This
means that imaging with Mo/YAG:Ce anode screen at 100 $\mu$A and 1
keV is equivalent to photographing the full moon on a clear night.
At lower currents, it becomes equivalent to taking images of a
moonless clear sky with stars.

We note here that the best practice to take pictures in the
described arrangement is as follows. The ISO has to be
1,000-3,000, because it drastically reduces pattern/banding noise
and electronic noise, and allows better discrimination of small
lighting features above the background when collecting for many
seconds. Secondly, top level cameras preserve the highest dynamic
range above 10 photographic stops, with the dynamic range measured
in photographic stops being calculated as $log_2\frac{full well
capacity [electrons]}{total noise [electrons]}$. This means that
high dynamic range and low noise at ISO of a few thousand enable
simultaneous detection of lighting features of exceptionally
different brightnesses, from $\sim$10 to $\sim$$10^5$ photons per
pixel. This is an extremely useful capability when one wants to
collect images through the entire I-V curve acquisition process,
from tens of nA to hundreds of $\mu$A.

\

\noindent \emph{Vacuum system}

In Fig.7, the vacuum diagram of the imager is presented.
Evacuation process of the vacuum chamber/vessel is three-fold. The
first pumping stage is an oil-free diaphragm pump (ultimate
pressure 1.5 Torr). Second stage pumping uses a turbo-molecular
pump: the pumping speed for N$_2$ is 67 l/s, the compression ratio
for N$_2$ and Ar is 10$^{11}$, the compression ratio for H$_2$ is
10$^5$, and the typical ultimate pressure with no high temperature
baking is 2$\times$$10^{-8}$ Torr at the backing pressure of 1.5
Torr. The third stage is an ion pump (pumping speed for N$_2$ is
30 l/s) that is turned on at 5$\times$$10^{-6}$ Torr which
significantly improves the pumping speed toward 2$\times$$10^{-8}$
Torr.

To quickly (in about 24 hours) achieve the nominal working
pressure of $\approx$3$\times$$10^{-9}$ Torr, which is
satisfactory enough to perform measurements, the following
procedure is established. The vacuum system does not have any
single Viton or rubber O-ring connections. All vacuum components
such as mechanical holders and screws are ultrasonically cleaned.
Step one is to clean in a mixture of distilled water and an acid
based detergent performed for about 30 minutes. Step two is to
clean in a 50/50 mixture of ethanol and acetone performed for
another 30 minutes. All parts are dried using dry nitrogen before
installation. All parts undergo a high temperature drying during
the first bake out of the system. The routine bake out procedure
is as follows. The system is heated to up to 140-145 $^\circ$C
when a pressure of 8$\times$$10^{-8}$ Torr or lower is achieved
through the described three-fold procedure. The choice of the bake
out temperature is limited by the Pfeiffer full range pirani/cold
cathode gauge in use (recommended baking temperature must not
exceed 150 $^\circ$C with magnet detached). The chamber is kept at
140-145 $^\circ$C for about 12 hours. The measure of a successful
baking procedure is if the pressure (with gauge disconnected,
calculated via the tabulated "ion current versus pressure" curve
provided for the ion pump by the manufacturer) returns to
8$\times$$10^{-8}$ Torr or better at 140-145 $^\circ$C. After
that, the heating is turned off and the chamber naturally cools
downs for another 12 hours and reaches the targeted
$\approx$3$\times$$10^{-9}$ Torr. This procedure is fully
automated. The final step is to close an isolation gate valve (the
one atop the turbo pump symbol in Fig.7) and turn off the
diaphragm and turbo pumps and wait until the pressure stabilizes
at $\approx3\times10^{-9}$ Torr with only the ion pump on.
Mechanical vibration free pumping is necessary because field
emission systems are sensitive to vibrations resulting in
cathode-anode gap fluctuations producing noisy current readings.
An SRS residual gas analyzer monitors gas composition in the
system and checks for any virtual or actual leaks.

\begin{figure}[t]
\includegraphics[width=8cm]{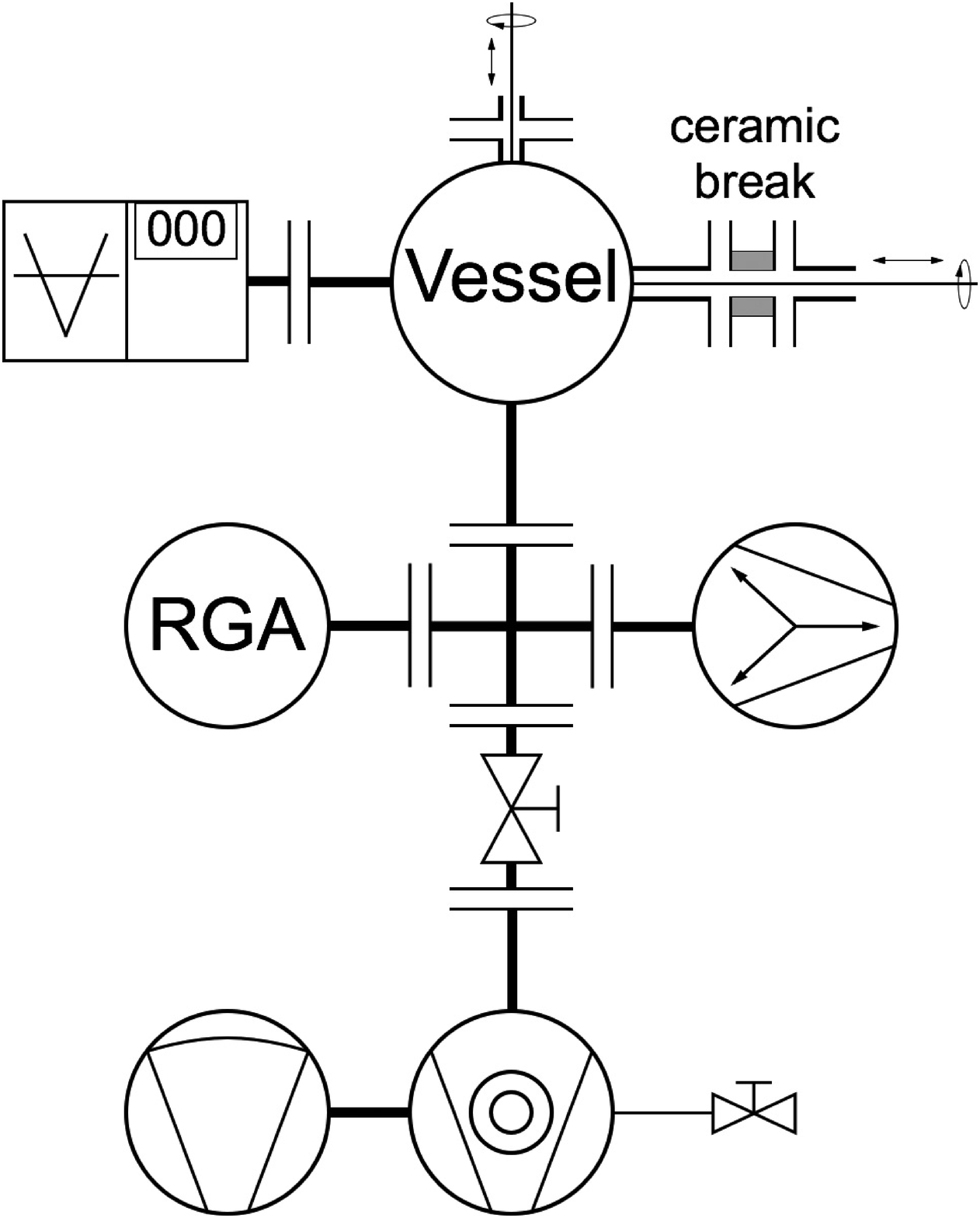}

\caption{Vacuum diagram of the imager}
\end{figure}

\

\noindent \emph{I-V curve measurement and imaging algorithm}

The entire data acquisition process is automated using the LabVIEW
programming environment. The Keithley 2410 simultaneously acts as
a voltage source (active mode: control signal in/feedback voltage
reading out) and a pico- to micro-ammeter (passive mode: current
reading out). The vacuum reading is taken directly from a Pfeiffer
controller connected to the full range pirani/cold cathode gauge.
Ion pump current/pressure is acquired using a DAQ USB-6003 by
National Instruments. Electron emission images are taken by the
Canon 6D equipped with a 50 mm lens and close-up lenses and
extensions; the standard work distance is set at 6 cm.

Fig.8 illustrates the algorithm to measure current as voltage is
swept. After the voltage is increased/decreased and before the
current measurement starts, there is a 1 second delay (typically
$dt_1$=1 s, but can be varied down to 0 s) introduced to allow
time for the electrometer internal circuit to settle. After the
$dt_1$ delay, the current is sampled $n$ times with a sampling
rate 1/$dt_2$. Typically $n$=20 and $dt_2$=300 ms, but both values
can be varied. The current, voltage, and pressure are sampled and
processed to obtain the mean value, the standard deviation and the
total measurement error. All are calculated online using
well-known statistics formulas

\[
\bar{x}=\frac{1}{n}\sum_{i=1}^n x_i \text{, (4)}
\]

\[
s=\sqrt{\frac{1}{n-1}\sum_{i=1}^n (x_i-\bar{x})^2} \text{, (5)}
\]

\[
\Delta x=t_{p,n}\cdot\frac{s}{\sqrt{n}}+\theta \text{, (6)}
\]
here $t_{p,n}$=2.09 is the Student's $t$-distribution coefficient
for a confidence interval of 95\%, $n$=20 is the number of
measurements at a given voltage, $\theta$=50 pA and 10 $\mu$V are
the ultimate accuracies of the Keithley 2410 electrometer for
current and voltage, respectively. All data, raw and processed,
are recorded and stored into files.

\begin{figure}[t]
\includegraphics[width=8cm]{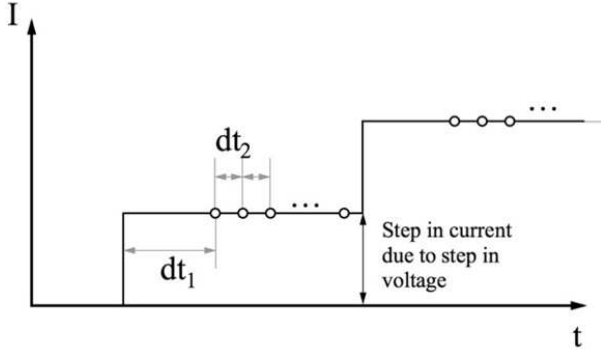}
\caption{Schematic diagram of current measurement algorithm}
\end{figure}

The software is designed such that I-V curve measurement and
collection of images using Mo/YAG:Ce or ITO/BASG anode screens are
synced within a specified run, with the voltage ramped up and
down. The entire block diagram is shown in Fig.9. The LabVIEW
program allows specification of a voltage step to repeatedly
acquire images with the Canon 6D as the voltage is swept up or
down. One can specify images to be taken every 20, 50, 100 V, etc.
while the voltage is being ramped. The voltage step size can be
between 1 V and 1 kV.

\begin{figure}[t]
\includegraphics[width=8cm]{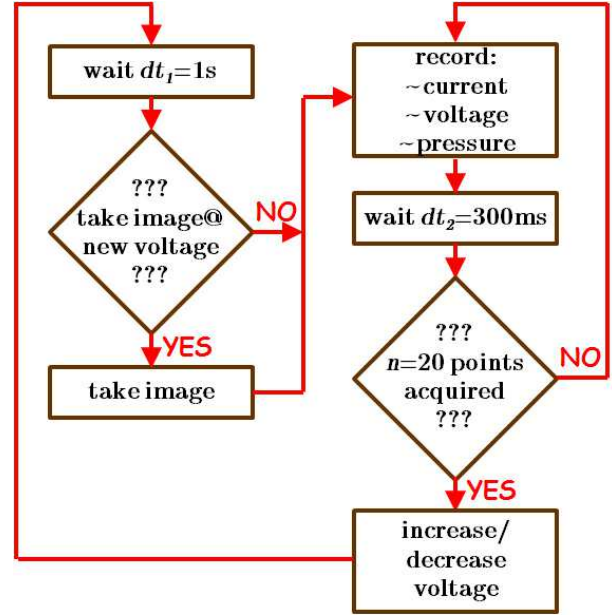}

\caption{Block diagram of the LabVIEW software for fully automatic
data and image acquisition}
\end{figure}

\

\noindent \textbf{Conclusions}

We presented a design and detailed description of an apparatus for
direct imaging of the electron field emission site distribution on
cathodes surfaces. We implemented a projection approach to the
imaging problem, making use of anode screens in the standard
parallel plate configuration. Detailed characterization and
performance metrics of the imaging Mo/YAG:Ce anode screen are
given, and recommendations for metal coating procedures are
presented based on our experience. In particular, we have
calculated electron to photon conversion efficiency as well as
measured the response and sensitivity of the optical system to the
emitted current. We also estimated that the lateral spatial
resolution of the imager can be on the order of 1-10 $\mu$m. As a
part of the imaging apparatus, the fully automated vacuum
control/monitor and data acquisition software was described.

The commissioned electron emission projection imager, in
combination with microscopy and spectroscopic methods, may unveil
electron emission induced processes on the micron and submicron
scale.

\

\noindent \textbf{Acknowledgment}

Euclid was supported by the Office of Nuclear Physics of DOE
through a Small Business Innovative Research grant \# DE-SC
0013145. We also thank

Alexander Zinovev of Materials Science Division at ANL for
providing the access to SARISA, a mass spectrometer. Use of SARISA
was supported by the U.S. Department of Energy, Office of Science,
Materials Sciences and Engineering Division.

Kiran Kovi and Anirudha Sumant of the Center of Nanoscale
Materials at ANL for providing us with the CNT sample. Use of the
Center for Nanoscale Materials, an Office of Science user
facility, was supported by the U.S. Department of Energy, Office
of Science, Office of Basic Energy Sciences, under Contract No.
DE-AC02-06CH11357.

Jiaqi Qiu and Richard Konecny of Euclid TechLabs for their kind
help with some of hardware manufacture.

SEM measurements were conducted in the Electron Microscopy Center
of the Center for Nanoscale Materials at Argonne National
Laboratory. Use of the Center for Nanoscale Materials, an Office
of Science user facility, was supported by the U.S. Department of
Energy, Office of Science, Office of Basic Energy Sciences, under
Contract No. DE-AC02-06CH11357.

\end{document}